\documentclass[conference]{IEEEtran}
\IEEEoverridecommandlockouts
\usepackage[linesnumbered,algoruled,boxed,lined,ruled, vlined]{algorithm2e}
\usepackage{cite}
\usepackage{float}
\usepackage{amsmath,amssymb,amsfonts}
\usepackage{breqn}
\usepackage{multirow}
\usepackage{graphicx}
\usepackage{textcomp}
\usepackage{xcolor}
\usepackage{hyperref}
\hypersetup{
   colorlinks=true,
    }

\def\BibTeX{{\rm B\kern-.05em{\sc i\kern-.025em b}\kern-.08em
    T\kern-.1667em\lower.7ex\hbox{E}\kern-.125emX}}
\begin{document}

\title{A Cost Effective Reliability Aware Scheduler for Task Graphs in Multi-Cloud System \\
}

\author{\IEEEauthorblockN{ Atharva Tekawade}
\IEEEauthorblockA{\textit{Department of Computer Science and Engineering,} \\
\textit{Indian Institute of Technology Jammu,}\\
Jammu, India \\
\texttt{2018uee0137@iitjammu.ac.in }}
\and
\IEEEauthorblockN{Suman Banerjee}
\IEEEauthorblockA{\textit{Department of Computer Science and Engineering} \\
\textit{Indian Institute of Technology Jammu,}\\
Jammu, India \\
\texttt{suman.banerjee@iitjammu.ac.in}}
}

\maketitle

\begin{abstract}
Many scientific workflows can be represented by a Directed Acyclic Graph (DAG) where each node represents a task, and there will be a directed edge between two tasks if and only if there is a dependency relationship between the two \emph{i.e.} the second one can not be started unless the first one is finished. Due to the increasing computational requirements of these workflows, they are deployed on cloud computing systems. Scheduling of workflows on such systems to achieve certain goals(e.g. minimization of makespan, cost, or maximization of reliability, etc.) remains an active area of research. In this paper, we propose a scheduling algorithm for allocating the nodes of our task graph in a heterogeneous multi-cloud system. The proposed scheduler considers many practical concerns such as pricing mechanisms, discounting schemes, and reliability analysis for task execution. This is a list-based heuristic that allocates tasks based on the expected times for which VMs need to be rented for them. We have analyzed the proposed approach to understand its time requirement. We perform a large number of experiments with real-world workflows: FFT, Ligo, Epigenomics, and Random workflows and observe that the proposed scheduler outperforms the state-of-art approaches up to 12\%, 11\%, and 1.1\% in terms of cost, makespan, and reliability, respectively.

\end{abstract}


\section{Introduction}
In recent times, \emph{cloud computing} has emerged as an alternative computing framework and become popular due to many features including 'pay as you use' kind of billing strategy, virtualization, rapid elasticity, on-demand use, and so on \cite{mastelic2014cloud}. Hence, cloud computing has become prominent in the execution of scientific workflows. Nowadays many scientific workflows from different domains such as \emph{Epigenomics} in bio-informatics, \emph{Cybershake} in earthquake engineering, etc. have been successfully deployed on commercial clouds \cite{zhou2019minimizing,farid2020scheduling}. Recently, due to large-scale computational resource and diversity requirements, multiple cloud providers club together to form a larger infrastructure, and such framework is known as a multi-cloud system \cite{tang2021reliability}. In many cases, they are managed by a third party. Each one of them provides its own set of Virtual Machines (VMs) and billing mechanisms.
\par A multi-cloud system is highly heterogeneous with its respective hardware, software, and network infrastructure. Hence, it is prone to failure as well. As a consequence, reliability becomes an important issue. Also, a scientific workflow is defined as a set of tasks where there are several dependencies among tasks. So the failure of one task may lead to the failure of the entire workflow. For a successful execution of the workflow, the cloud infrastructure must provide highly reliable computing services. To improve the reliability of a scientific workflow, several techniques have been investigated including the inclusion of checkpoints, fault tolerance, scheduling of tasks, etc. \cite{han2018checkpointing,wen2016cost,zhou2016cloud}. Among all these techniques, fault tolerance and the scheduling algorithm are the two effective approaches to increasing reliability. In the fault tolerance mechanism, where if any VM fails then there are backup VMs from other cloud providers that take care of the execution of the failed VM's work.

The rest of the paper is organized as follows. Section \ref{Sec:RW} highlights the related research work and the main contributions we make in this paper. Section \ref{Sec:SM} describes the system's model and problem formulation. The proposed list-based heuristic solution has been described in Section \ref{Sec:PA}. Section \ref{Sec:Experiments} contains the experimental evaluation of the proposed solution methodology. Finally, Section \ref{Sec:CFA} concludes our study and gives future research directions.

\section{Related Work} \label{Sec:RW}
\par In the past few years, there has been an extensive study on multi-cloud systems \cite{li2015data,diaz2015supporting}. These studies are in diverse directions such as trust management in multi-cloud framework \cite{li2015data}, in adaptive setting \cite{trihinas2015monitoring} and many more. Several scheduling strategies have also been proposed for multi-cloud systems \cite{miraftabzadeh2016efficient,kang2018dynamic}. Recently, there are a few methodologies for scheduling workflows in the multi-cloud system that have been proposed \cite{al2019task,sooezi2015scheduling,li2018scheduling}. In general, in a scientific workflow, tasks have a very high level of data and control flow dependency with the precedence task(s). Generally, these tasks are executed on a multi-cloud system, and hence, reliability becomes an important issue. Recently, several works focus on the reliability issues \cite{han2018checkpointing,wen2016cost,di2016toward,zhou2016cloud}. The closest to our study are FR-MOS \cite{farid2020scheduling} and FCWS \cite{tang2021reliability} which take makespan, cost and reliability into consideration. Both of these approaches consider simple billing mechanisms without considering communication costs and discounting schemes. So, there is enough space to work on this problem.
\par In this paper, we propose a scheduling algorithm for scientific workflows in a multi-cloud system considering communication cost and discounting scheme. In particular, we make the following contributions in this paper:

\begin{itemize}
\item We propose a system model to study workflow scheduling in a multi-cloud system considering communication costs and different billing mechanisms.
 
\item We also perform the reliability analysis for task execution considering that the failure of the tasks is modeled using Poisson Distribution.
\item We propose a list-based heuristic solution for scheduling a scientific workflow in a multi-cloud system with its complexity analysis.
\item The proposed solution approach has been implemented with real-world and randomly generated scientific workflows and compared with the state-of-art approaches.

\end{itemize}    

\section{Systems Model \& Problem Formulation} \label{Sec:SM}
In this section, we describe the system's model and describe our problem formally. For any positive integer $n$, $[n]$ denotes the set $\{1, 2, \ldots, n\}$.
\subsection{Tasks and Workflow}
A task is a job that needs to be executed on the multi-cloud system. A \emph{scientific workflow} consists of multiple interrelated tasks and is modeled as a DAG $G(V, E)$, where $V(G)=\{v_i : i \in [n]\}$ is the set of $n$ tasks and $E(G) = \{(v_i, v_j) : v_i, v_j \in V(G)\}$ is the set of edges. Here an edge $(v_i, v_j)$ signifies that there is a precedence relationship between tasks $v_i$ and $v_j$ \emph{i.e.} the execution of $v_j$ can only be started after the execution of $v_i$ is finished and its output is transferred to $v_j$. Further, the weight of the edge denoted by $w(v_i, v_j)$ denotes the amount of data to be transferred. The computational resource requirement of $v_i$ is denoted by $w(v_i)$. Let $pred(v_i)$ denote the set of immediate \emph{predecessor task(s)} of $v_i$ \emph{i.e.} $pred(v_i)=\{v_j: (v_j,v_i) \in E(G)\}$. Similarly, $succ(v_i)$ denotes the set of immediate \emph{successor task(s)} of $v_i$ \emph{i.e.} $succ(v_i)=\{v_j : (v_i,v_j) \in E(G)\}$. To have the proper notion of a first and last task, we add two \emph{pseudo-tasks}: $v_{entry}$ and $v_{exit}$ having zero computational resource requirements. Further, we add a directed edge from $v_{entry}$ to every $v$ \emph{s.t.} $pred(v) = \emptyset$, and we add a directed edge from every $v$ \emph{s.t.} $succ(v) = \emptyset$ to $v_{exit}$. The weight of these edges is set to zero. For simplicity we assume, $v_1 = v_{entry}$ and $v_n = v_{exit}$.

\subsection{Multi-Cloud Providers}
Cloud providers provide their computation services in the form of virtual machines (VMs) \cite{stillwell2012virtual}, \cite{tang2021reliability}. We consider $m$ different cloud providers each providing its own set of VMs. Let, the $k$-th cloud provider offer a total of $m_k$ different VMs, and $VM(k, p)$ denotes the $p$-th type VM offered by the $k$-th cloud provider. In general, a VM is characterized by its computational resources \emph{i.e.} CPU, Memory, Disk, etc. However, for simplicity we consider only one dimension (CPU) as done in \cite{farid2020scheduling}, \cite{tang2021reliability}. Let, $w(VM(k, p))$ denotes the computational resource of $VM(k, p)$. We abbreviate computational units by CU. Let $T_{exec}[v_i, VM(k,p)]$ denote the execution time of the task $v_i$ on the $VM(k, p)$, and this can be roughly calculated using Equation No. \ref{eqn1}.
\begin{equation} \label{eqn1}
T_{exec}[v_i, VM(k,p)] = \frac{w(v_i)}{w(VM(k, p))}
\end{equation}
Let, $T_{s}[v_i, VM(k,p)]$ and $T_{f}[v_i, VM(k,p)]$ denote the start time and finish time, respectively of executing $v_i$ on $VM(k,p)$. In this study, we do not assume pre-emption. Hence, once a task starts it executes till completion on its scheduled VM, leading to Equation No. \ref{eqn2}.
\begin{equation} \label{eqn2}
T_{f}[v_i, VM(k,p)] = T_{s}[v_i, VM(k,p)] + T_{exec}[v_i, VM(k,p)]
\end{equation}

The makespan is defined as the total time required to execute the whole workflow, and this will happen when the last task finishes its execution. This leads us to Equation No. \ref{eqn3}.
\begin{equation} \label{eqn3}
\emph{makespan} = T_{f}[v_n, VM(k,p)].
\end{equation}
where, $v_n$ is assumed to be scheduled on $VM(k, p)$.

\subsection{Network}
Let $B_k$ denote the bandwidth for the VMs of the $k$-th cloud. Similarly, let $B_{k', k}$ denote the bandwidth connecting the $k^{'}$-th and $k$-th clouds. Communication time between two tasks $v_i$ and $v_j$ is denoted by $T_{comm}[v_i, v_j]$ and this depends on the amount of data to be transferred and the clouds where the tasks are hosted \cite{tang2021reliability}. This can be computed using Equation No. \ref{eqn4}.
\begin{equation} \label{eqn4}
T_{comm}[v_i, v_j] = \left\{ 
  \begin{array}{ c l }
  \frac{w(v_i, v_j)}{B_{k', k}} & \quad \textrm{if } k' \neq k \\
    \frac{w(v_i, v_j)}{B_k} & \quad \textrm{otherwise}
  \end{array}
\right.
\end{equation}
where, $v_i, v_j$ are assumed to be scheduled on the $k^{'}$-th and $k$-th clouds, respectively. \\
A task can start its execution only if it has received output(s) from all its predecessor task(s). Hence, the start time of the task $v_j$ can be expressed by Equation No. \ref{eqn6}.  

\begin{equation} \label{eqn6}
T_{s}[v_j, VM(k,p)] =  \max_{v_i \in pred(v_j)} T_{f}[v_i, VM(k^{'},p^{'})] + T_{comm}[v_i, v_j]
\end{equation}
where, $v_i$ is assumed to be scheduled on $VM(k^{'},p^{'})$.

The start time of $v_1$ is zero irrespective of the VM it is scheduled on.

\subsection{Costs}
 During the execution of the \emph{workflow}, multiple costs are involved, the major ones being described below.
 \begin{itemize}
 \item \textbf{Rent Cost}:  This is the cost corresponding to the period for which the VM is rented to execute a task. The rent time of the task $v_j$ when executed on $VM(k, p)$is denoted by $T_{rent}[v_j, VM(k, p)]$ and is given by Equation No. \ref{eqn8}. \cite{tang2021reliability}.
 
\begin{equation} \label{eqn8}
\scriptsize
T_{rent}[v_j, VM(k,p)] = \max_{v_i \in pred(v_j)} (T_f[v_j, VM(k,p)] - T_{f}[v_i, VM(k^{'},p^{'})]). 
\end{equation}

where, $v_i$ is assumed to be scheduled on $VM(k^{'},p^{'})$. \\
 Let the corresponding cost be denoted by $C_{rent}[v_j, VM(k,p)]$, which will be formulated in Section \ref{SubSec:Pric_Scheme}.
\item \textbf{Communication Cost}: This is the cost associated with the transfer of data between tasks. The communication cost depends on the amount of data to be transferred and the clouds where the tasks are hosted \cite{barrett2011learning}. Let $c_{k', k}$ denote the price per unit data to transfer between the $k^{'}$-th and $k$-th cloud. Then the cost for transferring data between $v_i$ and $v_j$ denoted by $C_{v_i, v_j}$, is given by Equation No. \ref{eqn9}.
\begin{equation} \label{eqn9} 
C_{v_i, v_j} = w(v_i, v_j) \cdot c_{k', k}
\end{equation}
where, $v_i, v_j$ are assumed to be scheduled on the $k^{'}$-th and $k$-th clouds, respectively. \\
The total cost comprising of both rent and communication components for scheduling $v_j$ on $VM(k, p)$ is denoted by $C_{total}[v_j, VM(k, p)]$ and is given by Equation No. \ref{eqn10}.
\begin{equation} \label{eqn10} 
\scriptsize
C_{total}[v_j, VM(k, p)] = C_{rent}[v_j, VM(k, p)] + \sum_{v_i \in pred(v_j)} C_{v_i, v_j}
\end{equation}
\end{itemize}

\subsection{Pricing schemes} \label{SubSec:Pric_Scheme}
Most cloud providers divide their instance types into three major categories: on-demand, reserved and spot instances \cite{khatua2013novel}. In this study, we do not consider reserved instances because they assume that the user has already made a commitment agreement which may not always be true. We also do not consider spot instances because these types of VMs are quite unreliable \cite{monge2020cmi}. We now have a look at the basic pricing mechanisms employed by the different cloud providers for on-demand instances. For this, assume that $c_{k, p}$ denotes the on-demand cost per hour for renting $VM(k, p)$. \\
\emph{Microsoft Azure} follows a fine-grained scheme where the customer is charged per minute of usage \cite{farid2020scheduling}.
\begin{equation} \label{eqn11}
C_{rent}[v_i, VM(k,p)] = T_{rent}[v_i, VM(k,p)] \cdot \frac{c_{k,p}}{T_{minute}}
\end{equation}
\emph{Amazon Web Services (AWS)} bills the customer per hour of usage \cite{farid2020scheduling}. Hence, tasks can share billing periods without incurring additional costs \cite{cai2014heuristics}.
\begin{equation} \label{eqn12} 
C_{rent}[v_i, VM(k,p)] = \left\{ 
  \begin{array}{ c l }
  	\lfloor \frac{T_{rent}[v_i, VM(k,p)]}{T_{minute}} \rfloor \cdot c_{k,p} \\
    \lceil \frac{T_{rent}[v_i, VM(k,p)]}{T_{minute}} \rceil \cdot c_{k,p} 
  \end{array}
\right.
\end{equation}
In the above equations, $T_{minute} = 60$ and $T_{rent}[v_i, VM(k,p)]$ is assumed to be converted to the nearest greatest minute. \\ 
\emph{Google Cloud Platform (GCP)} follows a resource-based pricing mechanism, where the price per unit resource depends on the machine family \cite{gcppricing}. Let $[\phi_i], i \in [r]$ denote the various machine families, and $c_i$ denote the cost per unit resource per unit hour for that family. Each VM is defined by a fixed number of compute and memory units. As mentioned earlier, we only consider the compute units in this study. Hence, the pricing for $VM(k,p) \in \phi_i$, can be given by:
\begin{equation} \label{eqn 13}
c_{k,p} = w(VM(k,p)) \cdot c_i
\end{equation}
Like \emph{Azure}, \emph{GCP} follows a per-minute billing policy given by Equation No. \ref{eqn11} \cite{gcppricing}. 

%

\begin{table}
		\centering
		\begin{tabular}{|c | c | c|} 
			\hline
			S.No. & Usage level & Price per incremental time unit \\ [0.3ex] 
			\hline\hline
			1. & $0\% - 25\%$ & 100\% \\ 
			\hline
			2. & $25\% - 50\%$ & 80\% \\
			\hline
			3. & $50\% - 75\%$ & 60\% \\
			\hline
			4. & $75\% - 100\%$ & 40\% \\
			\hline
		\end{tabular}
		\vspace{0.1 cm}
		\caption{Discounts for n1-type machines}
		\label{tab2}
	\end{table}

\subsection{Discount schemes}
In this section, we discuss Sustained Use Discounts, a discounting scheme offered by \emph{GCP}. When resources are being used for a sustained amount of time, discounts will be offered depending on the period for which they were running. When multiple VM instances of a family are running together within a billing period, they are first broken down into their unit components and then combined into multiple non-overlapping instances each with the longest possible duration. Since the price per unit resource is the same and the discount increases with duration, this process of combining the VMs resource-wise into the longest possible durations results in the largest possible discount, which is what Google compute engine does automatically \cite{susdisc}. At the end of the billing period, the VMs originally purchased are combined to produce new VMs with new rent times denoted by $VM(k^{'}, p^{'})$ and $T_{rent}[VM(k^{'},p^{'})]$, respectively. However, they follow the same pricing $c_i$ according to the original family. The new cost denoted by $C_{rent}[VM(k^{'},p^{'})]$ can be found using Equation No. \ref{eqn11} with a discount applied on top of it. The amount of discount depends on $T_{rent}[VM(k^{'},p^{'})]$ and the discounting scheme followed by that machine type, Table \ref{tab2} shows an illustration of the discounting scheme followed by n1-type machines \cite{susdisc}.


\begin{figure}[!ht]
\centering
\includegraphics[scale=0.2]{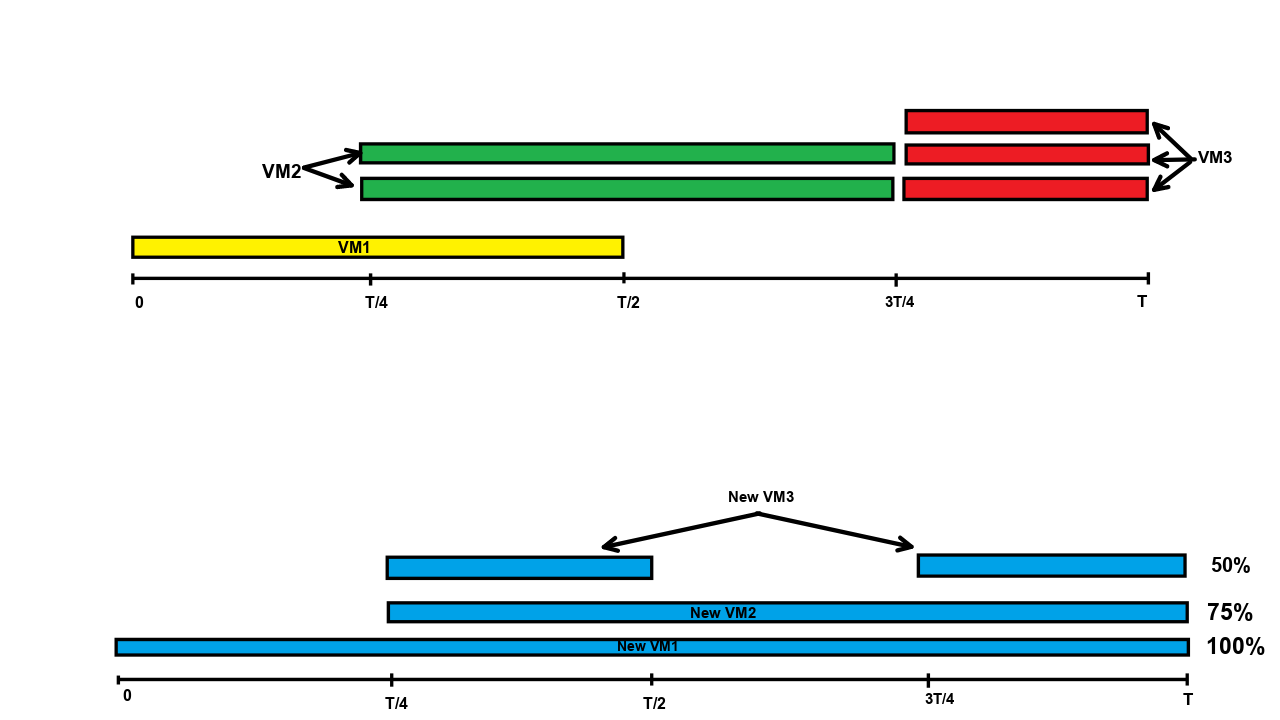}
\caption{Combining resources for sustained discount}
\label{Fig:sustained_discount}
\end{figure} 

Fig.\ref{Fig:sustained_discount} shows an illustration of how the instances are combined to achieve maximum possible sustained use discounts. From the example, we see that from 0 to T/2, an instance with 1 CU is running. From T/4 to 3T/4, an instance with 2 CU is running and from 3T/4 to T an instance with 3 CU is running, where T denotes the billing period. Since, the greatest common divider of $1$, $2$, and $3$ is 1, we split each VM into parts of 1 CU each. Then, following the principle of combining resources into the longest possible durations, we get the following new instance types: 1 CU for 100\% of the billing period, 1 CU for 75\% of the billing period, and 1 CU for 50\% of the billing period as shown in Fig.\ref{Fig:sustained_discount}. The algorithm for obtaining the longest length intervals is named as \textbf{M}aximal \textbf{L}ength \textbf{I}ntervals (abbreviated as \textbf{MLI}) and is presented in Algorithm 1. \\
We now proceed to formulate the overall cost incurred for scheduling the workflow. Let $cost_1$ denotes the cost incurred due to \emph{Azure} and \emph{AWS} clouds, while $cost_2$ denotes the cost incurred due to \emph{GCP} cloud. The overall cost denoted by $cost$, includes $cost_1, cost_2$ and the communication costs.

\begin{equation} \label{eqn14}
cost_1 = \sum_{i=1}^{n} \sum_{k=1}^{m} \sum_{p=1}^{m_k} \textbf{X}_{i,k,p} \cdot C_{rent}[v_i, VM(k,p)]
\end{equation}
where, the $k$-th cloud is of \emph{Azure} or \emph{AWS} type and \textbf{X} denotes the task allocation matrix defined below: \\.

$\textbf{X}_{i, k, p} = \left\{ 
  \begin{array}{ c l }
  1 & \quad \textrm{if $v_i$ is scheduled on $VM(k, p)$} \\
    0 & \quad \textrm{otherwise}
  \end{array}
\right.$

\begin{equation} \label{eqn15}
cost_2 = \sum C_{rent}[VM(k^{'},p^{'})]
\end{equation}
\begin{equation} \label{eqn16}
cost = cost_1 + cost_2 + \sum_{j=1}^{n} \sum_{v_i \in pred(v_j)} C_{v_i, v_j}
\end{equation}

\subsection{Reliability Analysis}
Reliability is defined as the probability of a failure-free execution of the workflow. The occurrence of a failure is modeled as a Poisson Distribution \cite{farid2020scheduling}. Let $\lambda_{k, p}$ denote the parameter of the distribution for $VM(k, p)$. Similarly, let $\lambda_{k}$ and $\lambda_{k', k}$ denote the parameters for the communication links for the $k$-th cloud and for those between the $k^{'}$-th and the $k$-th cloud, respectively. The probability that the task $v_j$ will execute successfully on $VM(k, p)$ is denoted by $\mathcal{R}[v_j, VM(k, p)]$ and is given by Equation No. \ref{eqn17}. 
\begin{equation} \label{eqn17}
\scriptsize
\mathcal{R}[v_j, VM(k, p)] = \prod_{v_i \in pred(v_j)} \mathcal{R}[v_i, v_j] \cdot e^{-\lambda_{k, p} \cdot T_{rent}[v_j, VM(k,p)]}
\end{equation}
where, $\mathcal{R}[v_i, v_j]$ denotes the reliability of the communication link between $v_i$ and $v_j$ given by the below equation:
\begin{equation} \label{eqn18}
\scriptsize
\mathcal{R}[v_i, v_j] = \left\{ 
  \begin{array}{ c l }
   e^{-\lambda_{k', k} \cdot T_{comm}[v_i, v_j]} & \quad \textrm{if } k' \neq k \\
   e^{-\lambda_{k} \cdot T_{comm}[v_i, v_j]} & \quad \textrm{otherwise}
  \end{array}
\right.
\end{equation}
where $v_j$ is assumed to be scheduled on the $k^{'}$-th cloud. \\
The workflow will execute successfully if all the tasks execute successfully. Assuming the failures are independent, the reliability of the workflow is given by the product of the reliability of each task which has been expressed using Equation No. \ref{eqn19}.
\begin{equation} \label{eqn19}
reliability = \prod_{j=1}^{n} \mathcal{R}[v_j, VM(k, p)]
\end{equation}

\subsection{Problem Formulation}
Now we state the problem that we have worked out in this paper. We wish to allocate the tasks to VMs that minimizes cost, makespan and maximizes reliability of the workflow. The problem can now be mathematically  stated as follows: \\
\textbf{Minimize} $\left\{ 
  \begin{array}{ c l }
   	 \textit{makespan} \\
   	 \textit{cost}
  \end{array}
\right.$
\textbf{Maximize} $\left\{ 
  \begin{array}{ c l }
  	 \textit{reliability}
  \end{array}
\right.$ \\ \\

\section{Proposed Algorithm} \label{Sec:PA}
In this section we describe our algorithm  \textbf{R}ent \textbf{D}ependent \textbf{C}ost \textbf{E}ffective \textbf{W}orkflow \textbf{S}cheduling (abbreviated as \textbf{RDCWS}), a list-based approach having two phases. In the first phase, tasks are ordered by computing their ranks. In the second phase, tasks are mapped to VMs one by one in decreasing order of their ranks.
\subsection{Task Ordering}
Let $\overline{TE(v_j)}$ and $\overline{TC(v_j)}$ denote the median execution and communication times for the task $v_j$ defined below:
\begin{equation} \label{eqn20}
    \overline{TE(v_j)} = \frac{w(v_i)}{\overline{w(VM(k, p))}}
\end{equation}
\begin{equation} \label{eqn21}
    \overline{TC(v_j)} = \left\{ 
  \begin{array}{ c l }
  \frac{\max_{v_i \in pred(v_j)}w(v_i, v_j)}{\overline{B}}  & \quad \textrm{if } v_j \neq v_1 \\
    0 & \quad \textrm{otherwise}
  \end{array}
\right.
\end{equation}
where $\overline{w(VM(k, p))}, \overline{B}$ denote the median VM computation capacity and bandwidth, respectively.
The rank of task $v_i$ denoted by $Rank(v_i)$ is calculated as follows:

\begin{equation} \label{eqn22}
\scriptsize
Rank(v_i)=
\left\{ 
  \begin{array}{ c l }
  \overline{TE(v_i)} + \overline{TC(v_i)}  & \quad \textrm{if } v_i = v_n \\
 \underset{v_j \in succ(v_i)}{max}    \{Rank(v_j)\} +  \overline{TE(v_i)} +   \overline{TC(v_i)} & \quad \textrm{otherwise}
  \end{array}
\right.
\end{equation}

The order is obtained by sorting tasks in decreasing order of their ranks. 

\subsection{VM Allocation}
In this section, we describe the VM Allocation strategy that aims to assign tasks to VMs according to the above-computed order considering the cost, makespan, and reliability objectives. Let us consider the allocation of task $v_j$. For costs, observe that if $v_j$ is assigned to a VM of \emph{GCP} cloud, then sustained discounts may apply and costs of previously allocated tasks might change. Hence, while allocating it, we need to consider the total cost denoted by $C[v_j, VM(k, p)]$, which can be obtained by considering Equation No. \ref{eqn14} - \ref{eqn16} up to node $v_j$ instead of $v_n$ (which instead corresponds to the whole task graph). As for makespan and reliability, we only consider the finish time and reliability respectively of $v_j$ alone when executed on $VM(k, p)$. From Equation No. \ref{eqn11}, \ref{eqn12} and \ref{eqn17}, we can see that lesser value of rent time is favorable for both cost and reliability. Hence for tasks with higher expected rent times, it is better to choose a VM that minimizes cost and maximizes reliability. Below, we get an estimate of the rent time of a task.

$T_{rent}[v_j, VM(k, p)] = \max_{v_i \in pred(v_j)}(T_f[v_j, VM(k, p)] - T_f[v_i, VM(k', p')]) \text{ \ldots \ldots From Eq. \ref{eqn8}} \\ = \max_{v_i \in pred(v_j)}(T_s[v_j, VM(k, p)] + T_{exec}[v_j, VM(k, p)] - T_f[v_i, VM(k', p')]) \text{ \ldots \ldots From Eq. \ref{eqn2}} \\ = T_s[v_j, VM(k, p)] - \min_{v_i \in pred(v_j)}T_f[v_i, VM(k', p')] + T_{exec}[v_j, VM(k, p)]  = \max_{v_i \in pred(v_j)}(T_f[v_i, VM(k', p')]+T_{comm}[v_i, v_j])- \min_{v_i \in pred(v_j)}T_f[v_i, VM(k', p')] + T_{exec}[v_j, VM(k, p)] \text{ \ldots \ldots From Eq. \ref{eqn6}} \\
\geq \max_{v_i \in pred(v_j)} T_{comm}[v_i, v_j] + T_{exec}[v_j, VM(k, p)]$. \\


The above obtained lower bound is used as an approximation for the rent time of a task denoted by $\tilde T_{rent}[v_j, VM(k, p)]$ calculated as shown below:

\begin{equation} \label{eqn23}
\tilde T_{rent}[v_j, VM(k, p)] = \overline{TC(v_j)} + \overline{TE(v_j)}
\end{equation}

We sort the tasks in decreasing order of their expected rent times $\tilde T_{rent}[v_j, VM(k, p)]$ in a list $\tilde T_{rent}$. For the first $\ell(1 \leq \ell \leq n)$ tasks with the highest expected rent times, we allocate the task based on a normalized linear combination of only its cost with weight $\alpha_1$ and reliability with weight $\beta_1$. For the other tasks, we allocate them based on a normalized linear combination of cost with weight $\alpha_2$, reliability with weight $\beta_2$ and finish time (which relates to makespan) with weight $\gamma_2$, where normalization is done using min-max normalization. We name the metric involved in deciding the allocation as \emph{allocation metric} denoted by $AM[v_j, VM(k, p)]$ when $v_j$ is scheduled on \emph{VM(k, p)} and is subsequently defined below:

\begin{dmath} \label{eqn24} 
AM[v_j, VM(k, p)] = \left\{ 
  \begin{array}{ c l }
  	\alpha_1 \cdot \frac{C[v_j, VM(k, p)] - C_{\min}}{C_{\max} - C_{\min}} + \beta_1 \cdot \frac{\mathcal{R}_{\max} - \mathcal{R}[v_j, VM(k, p)]}{\mathcal{R}_{\max} - \mathcal{R}_{\min}} \\
    \alpha_2 \cdot \frac{C[v_j, VM(k, p)] - C_{\min}}{C_{\max} - C_{\min}} + \beta_2 \cdot \frac{\mathcal{R}_{\max} - \mathcal{R}[v_j, VM(k, p)]}{\mathcal{R}_{\max} - \mathcal{R}_{\min}} + \\ \gamma_2 \cdot \frac{T_f[v_j, VM(k, p)] - T_{\min}}{T_{\max} - T_{\min}}
  \end{array}
\right.
\end{dmath}

where, $\alpha_1 + \beta_1 = \alpha_2 + \beta_2 + \gamma_2 = 1$ and $C_{\min}, \mathcal{R}_{\min}, T_{\min}, C_{\max}, \mathcal{R}_{\max}, T_{\max}$ denote the minimum and maximum values of cost, reliability and finish times respectively when $v_j$  is executed over all VMs. Finally, $v_j$ is assigned to the VM with \emph{least allocation metric}. The final algorithm (RDCWS) is presented in Algorithm 2.

\begin{algorithm}[!htb]
\label{Algo:1}
\caption{MLI Algorithm}
	\KwIn{List of intervals $L$ characterized by their start and finish times}
	\KwOut{List of new intervals $L'$}
	Sort intervals according to increasing order of their start times, breaking ties arbitarily; \\
	$L' \longleftarrow []$\;
	\For{$interval \in L$} {
		$new\_interval \longleftarrow [interval]$\;
		$curr\_interval \longleftarrow interval$\;
		\For{$int \in L$} { \textcolor{blue}{\tcp{start adding intervals to current interval}}
			\If{$int.start\_time < curr\_interval.finish\_time < int.finish\_time$} {
			\textcolor{blue}{\tcp{add non-overlapping part of partially overlapping interval $int$}} \
			 	$temp \longleftarrow int.finish\_time$ \;
			 	$int.finish\_time \longleftarrow curr\_interval.finish\_time$ \;
			 	$curr\_interval.finish\_time \longleftarrow temp$; 
			}         
           	\If{$curr\_interval.fin \leq int.start$}{
           	\textcolor{blue}{\tcp{add completely non-overlapping interval $int$}} \
                 Add $int$ to $L^{'}$\;
                 $curr\_interval \longleftarrow int$\; \textcolor{blue}{\tcp{update current interval to the newly added interval}}
                 Remove $int$ from $L$ \; \textcolor{blue}{\tcp{Remove interval $int$ from $L$ because it is just added to $L^{'}$}}
			}
		}
		Add $new\_interval$ to $L^{'}$\; \textcolor{blue}{\tcp{add the newly obtained longest possible duration interval}}
		Remove $interval$ from $L$ \;
	}
	\KwRet $L'$
\end{algorithm}

\begin{algorithm}[!htb]
\caption{RDCWS Scheduling Algorithm}
	\KwIn{Task graph, Multi-cloud system parameters}
	\KwOut{Task allocation matrix \textbf{X}}
	Obtain task ordering in list $L$ using Equation No. \ref{eqn22}\;
	Sort tasks by decreasing order of expected rent times in list $\tilde T_{rent}$ using Equation No. \ref{eqn23}; \\
	$L' \longleftarrow []$; \textcolor{blue}{\tcp{List to keep spare VMs}}\
	\For{$v_j \in L$}{
		\For{$k \in [m], p \in [m_k]$ or $VM(k, p) \in L'$} {
			Calculate $T_f[v_j, VM(k,p)]$ using Equation No. \ref{eqn2}\;
			Calculate $C[v_j, VM(k,p)]$ according to Equation No. 					\ref{eqn14} - \ref{eqn16}\;
			Calculate $ R[v_j, VM(k,p)]$ according to Equation No. 							\ref{eqn17}\;
			Calculate $AM[v_j, VM(k,p)]$ according to Equation No.  \ref{eqn24}\;
		}
	}
	Choose the \emph{VM(k,p)} with \emph{least allocation metric}; \\
	Schedule $v_j$ on $VM(k,p)$; \\
	\If{$VM(k,p)$ has remaining time} {
		Add $VM(k, p)$ to $L'$\; 
	}
	$\textbf{X}_{j, k, p} \longleftarrow 1$; \\
	\KwRet \textbf{X}\;
	\label{Algo:2}
\end{algorithm}

\subsection{Analysis}
The overall complexity of our algorithm depends on the total number of VMs available: $\sum_{k=1}^{m} \sum_{p=1}^{m_k} 1 = \mathcal{N}$. As seen from Equation No. \ref{eqn22} and \ref{eqn23}, calculating the rank and approximate rent time for one task takes $\mathcal{O}(\mathcal{N} + n)$ time. Hence, the total time required for all the tasks is $\mathcal{O}(n \cdot (\mathcal{N} + n))$. Sorting the tasks based on rank and rent times takes $\mathcal{O}(n \cdot \log n)$ time. Now, for each task-VM pair, calculating finish times, reliabilities, etc. takes $\mathcal{O}(n)$ time as seen from Equation No. \ref{eqn2}, \ref{eqn6} and \ref{eqn17}. For costs, it is a bit more challenging. \emph{Azure} and \emph{AWS} costs can be calculated in a straightforward manner in $\mathcal{O}(n)$ time as seen from Equation No. \ref{eqn8}, \ref{eqn10}, \ref{eqn11}, \ref{eqn12}. For \emph{GCP}, the intervals need to be rearranged. If we have $t$ intervals at a time, it takes $\mathcal{O}(t^2)$ time to rearrange them as seen from Algorithm 1, where $t$ depends on the amount of \emph{GCP} VMs hosted till now: $t = \sum w(VM(k, p))$. In the worst case, $t = n \cdot w_{\max}$, where $w_{\max}$ is the largest size \emph{GCP} VM. Apart from new VMs, searching in the list $L'$ for an already launched VM takes $\mathcal{O}(n)$ time as atmost $n$ tasks are scheduled till now. Lastly, the parameter $\ell$ can be varied from $1$ to $n$ generating a family of solutions. Hence, the overall time complexity of RDCWS is $\mathcal{O}((\mathcal{N} + n) \cdot n^4 \cdot w_{\max}^2)$.

\subsection{Flexibility of RDCWS}
RDCWS can be tuned according to the use case as shown:
\begin{itemize}
\item Depending on which metric is more important, the weights can be adjusted accordingly. In this study, we give more importance to cost over makespan and reliability. The weights are thus chosen as: $\alpha_1 = 0.8, \beta_1 = 0.2$ and $\alpha_2 = 0.6, \beta_2 = 0.2, \gamma_2 = 0.2$. 
\item By varying the parameter $\ell$, RDCWS generates a family of solutions. Depending on our requirements, we can choose the appropriate solution. For example, by choosing a small value of $\ell$, we allocate more tasks based on finish time, hence improving makespan. Out of the family of solutions generated by RDCWS, we choose the one with minimum cost for comparison in our experiments.
\item By replacing $C[v_j, VM(k, p)]$ by $C_{total}[v_j, VM(k, p)]$ in Equation No. \ref{eqn24}, RDCWS can be used for scheduling when discounting schemes are not considered. In this case, we only consider the cost of scheduling $v_j$ alone as sustained use discounts don't apply and complexity is $\mathcal{O}((\mathcal{N} + n) \cdot n^2)$.
\end{itemize}

\section{Experimental Evaluation} \label{Sec:Experiments}
In this section, we describe the experimental evaluation of the proposed solution
approach. First, we start by describing the experimental setup and example task graphs.
\subsection{Experimental setup and Task graphs}
We implement the proposed solution approach on a workbench system with i5 $10^{th}$ generation processor and 32GB memory in Python 3.8.10. For this study, we consider three real world task graphs which are widely used in literature \cite{farid2020scheduling}, \cite{tang2021reliability}, \cite{xie2017energy} for comparison and randomly generated graphs for evaluation described below.
\begin{itemize}
    \item \emph{Fast Fourier transform(FFT):} The number of tasks $n = (2 + \rho) \cdot 2^\rho -1 $ \cite{topcuoglu2002performance}. We consider applications with $\rho = 4, 5$ or $n = 95, 223$.
    \item \emph{LIGO:} We consider applications with $n = 30, 100$\cite{szabo2012evolving}.
    \item \emph{Epigenomics:} We consider applications with $n = 24, 100$ \cite{szabo2012evolving}.
    \item \emph{Random:} As the name suggests, random graphs are generated in a random fashion. We consider applications with $n = 100, 200$.
\end{itemize}
As seen fom above, each task graph is considered for two node sizes: small (S) and large (L).
\subsection{Experimental description}
Each type of task graph is run for ten instances and the average values are reported. As mentioned earlier, RDCWS can be used with or without discounts, hence for each instance run, two experiments are performed: One considering discounting scheme and one without considering the discounting scheme. We consider 6 cloud vendors in our multi-cloud system: 2 from \emph{AWS}, 2 from \emph{Azure} and 2 from \emph{GCP}. VM computation capacities are chosen uniformly from $[1, 70]$ and prices are chosen approximately proportional to their computation capacity \cite{tang2021reliability}. For \emph{GCP}, we consider two machine families, one for each cloud provider with capacities and prices scaling proportionately \cite{gcppricing}. Additionally, discount schemes for both families are chosen according to Table \ref{tab2}. The billing period T is usually set to one-month \cite{susdisc}, but for experimental purposes, we set it to a much smaller value of ten hours. To model large data, task computation capacity and edge sizes (in Mb) of graph nodes are chosen uniformly at random between $[10, 10^5].$ The average communication bandwidth within a cloud is set to 100Mb/s and across clouds is set to 20Mb/s \cite{tang2021reliability}. The average cost for sending data within clouds of the same type is set to $0.01$, and across clouds of different types is set to $0.02$ per unit of data (in Gb). Within the same cloud, it is free. Lastly, the Poisson parameters are chosen uniformly from the ranges: $\lambda \in [10^{-9}, 10^{-8}]$.  Our algorithm is compared with the state-of-art algorithms: FR-MOS \cite{farid2020scheduling} and FCWS \cite{tang2021reliability} in terms of cost, makespan, and reliability. FCWS can be easily extended to include communication costs by replacing $C_{rent}[v_j, VM(k,p)]$ with $C_{total}[v_j, VM(k,p)]$ or $C[v_j, VM(k,p)]$ while allocation when discounts are not considered and considered respectively. 

\begin{figure*}[!ht]
\centering
\begin{tabular}{ccc}
\includegraphics[width=0.24\textwidth]{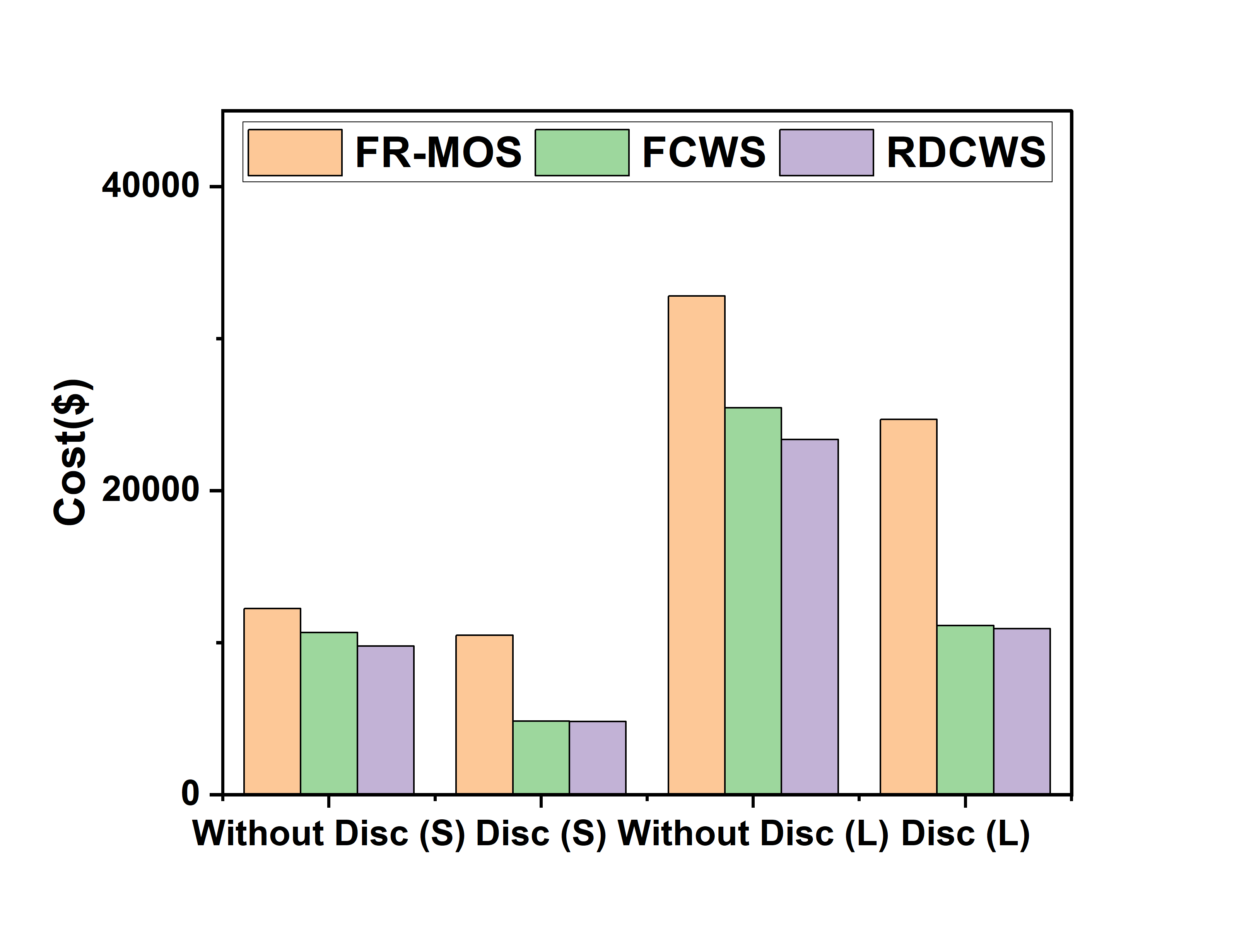} &
\includegraphics[width=0.24\textwidth]{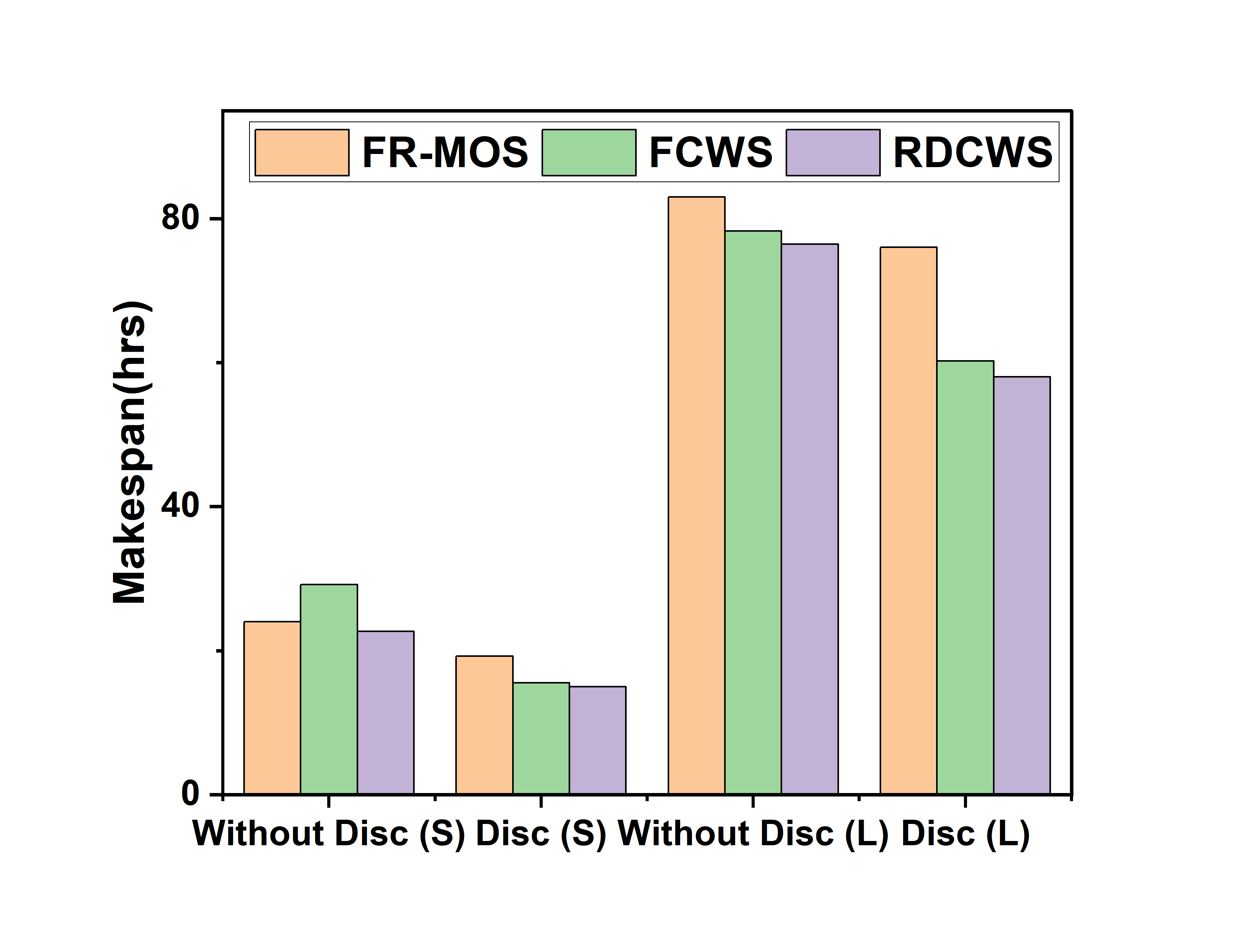} &
\includegraphics[width=0.24\textwidth]{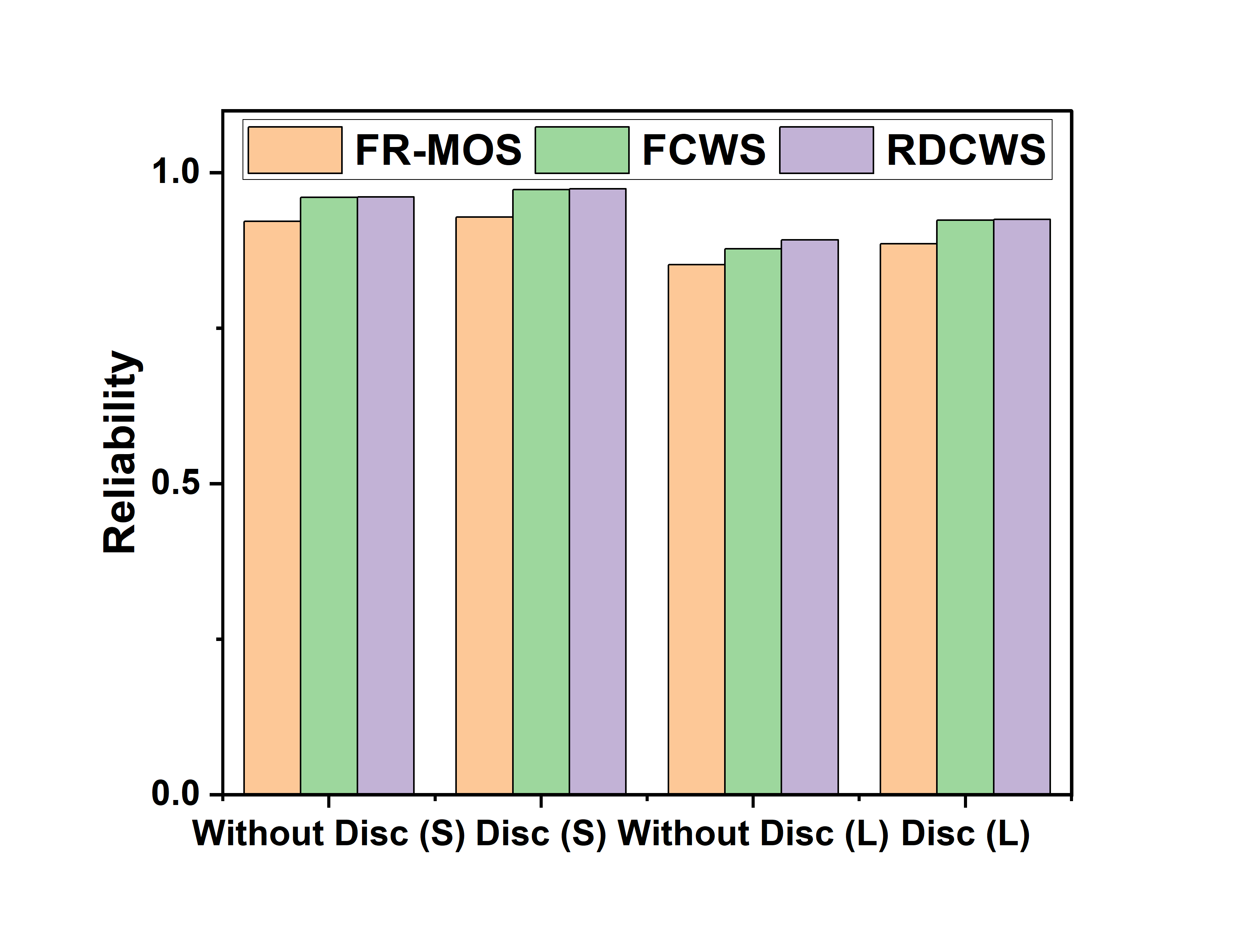}
\\
(a) & (b) & (c) \\
\end{tabular}
\caption{FFT applications with $\rho = 4, 5$}
\label{Fig2}
\end{figure*}

\begin{figure*}[!ht]
\centering
\begin{tabular}{ccc}
\includegraphics[width=0.24\textwidth]{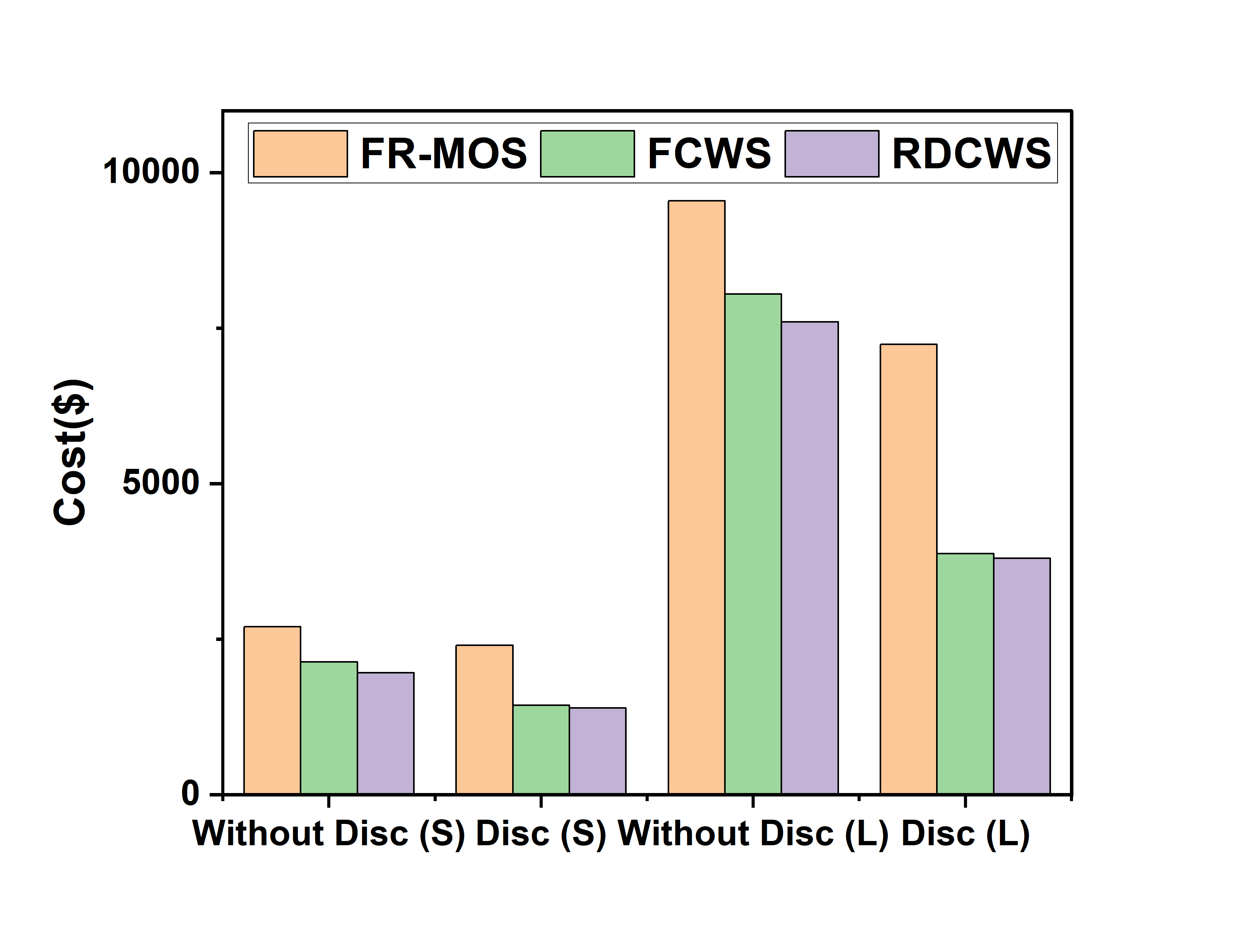} &
\includegraphics[width=0.24\textwidth]{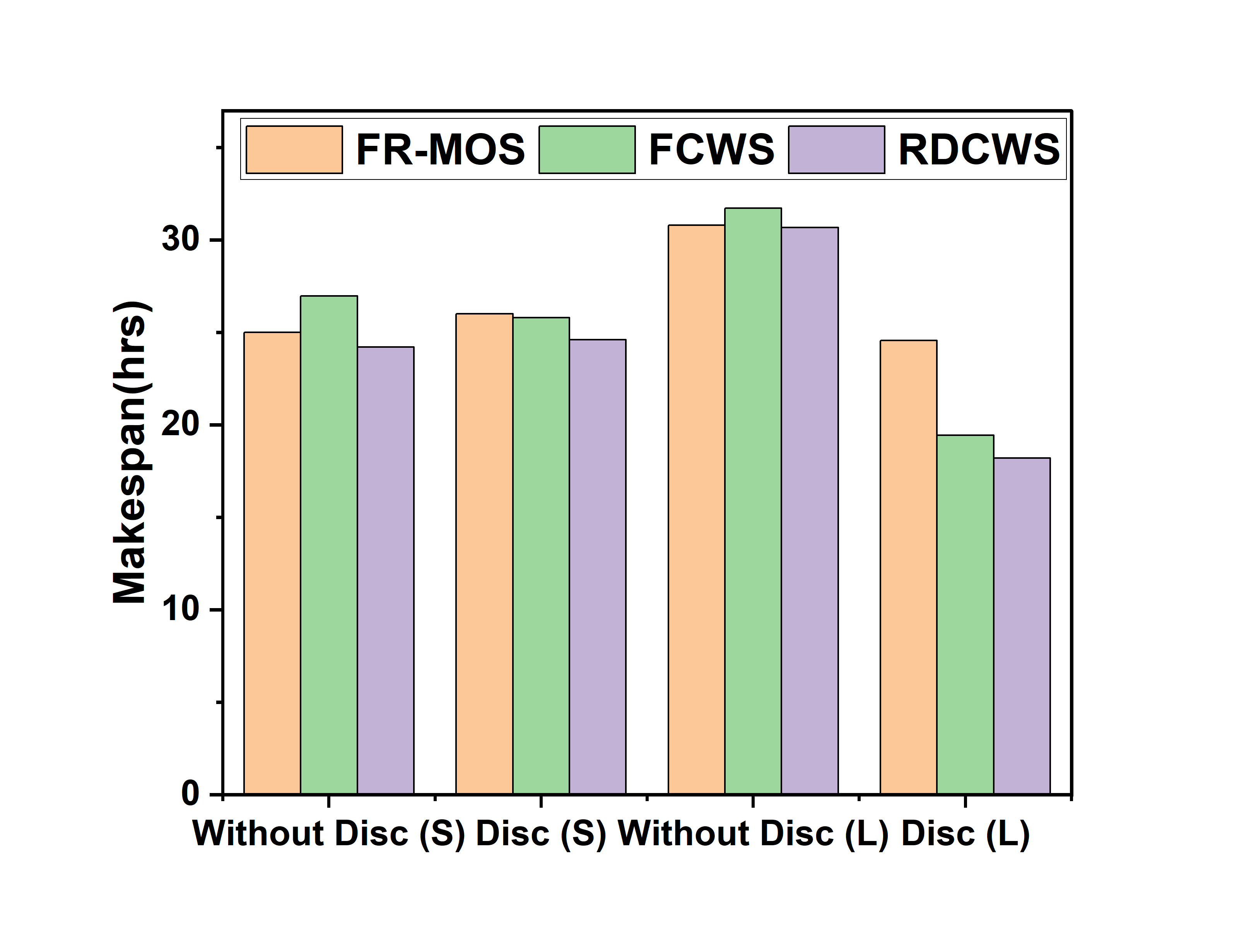} &
\includegraphics[width=0.24\textwidth]{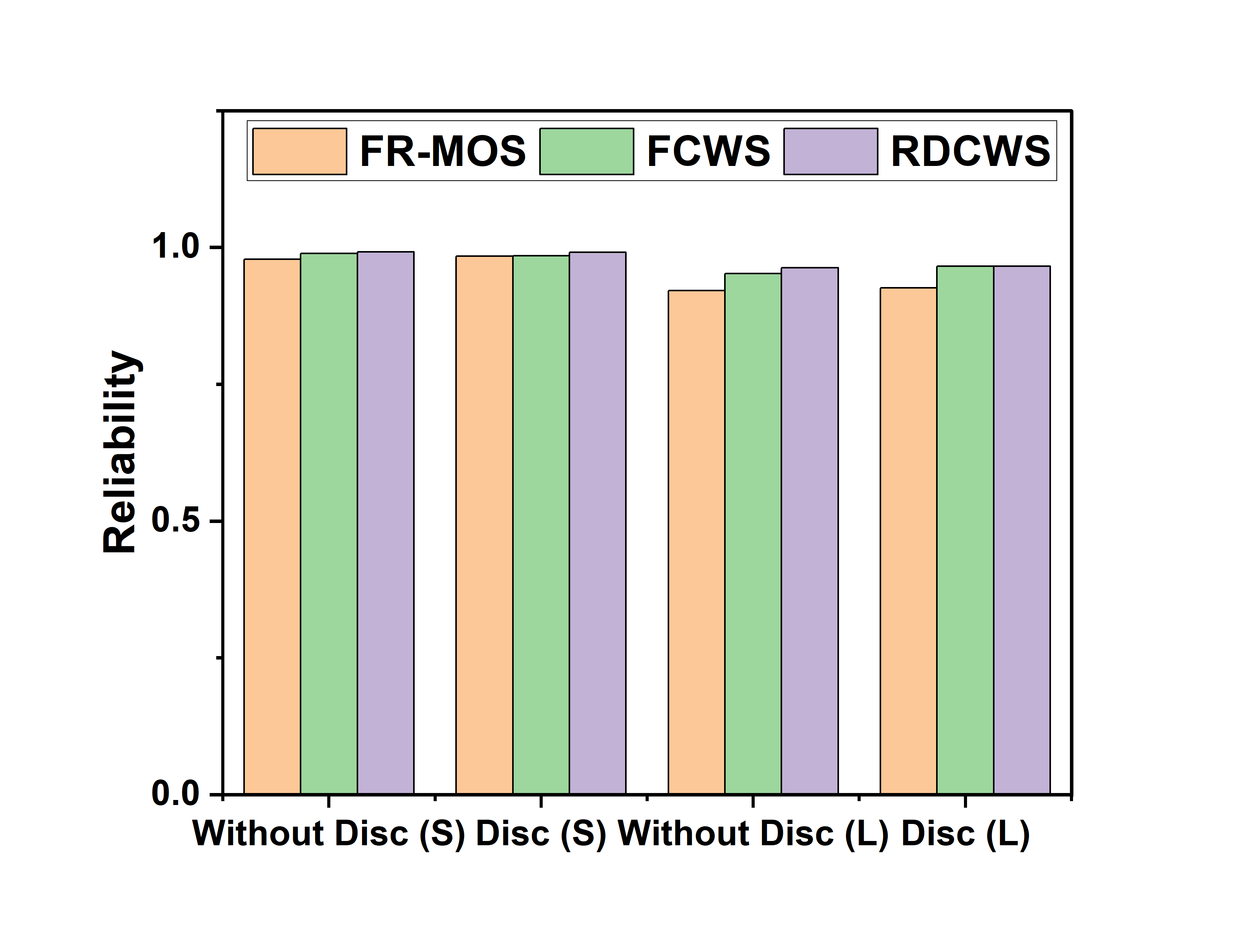} 
\\
(a) & (b) & (c) \\
\end{tabular}
\caption{Ligo applications with $n = 30, 100$}
\label{Fig3}
\end{figure*}

\begin{figure*}[!ht]
\centering
\begin{tabular}{ccc}
\includegraphics[width=0.24\textwidth]{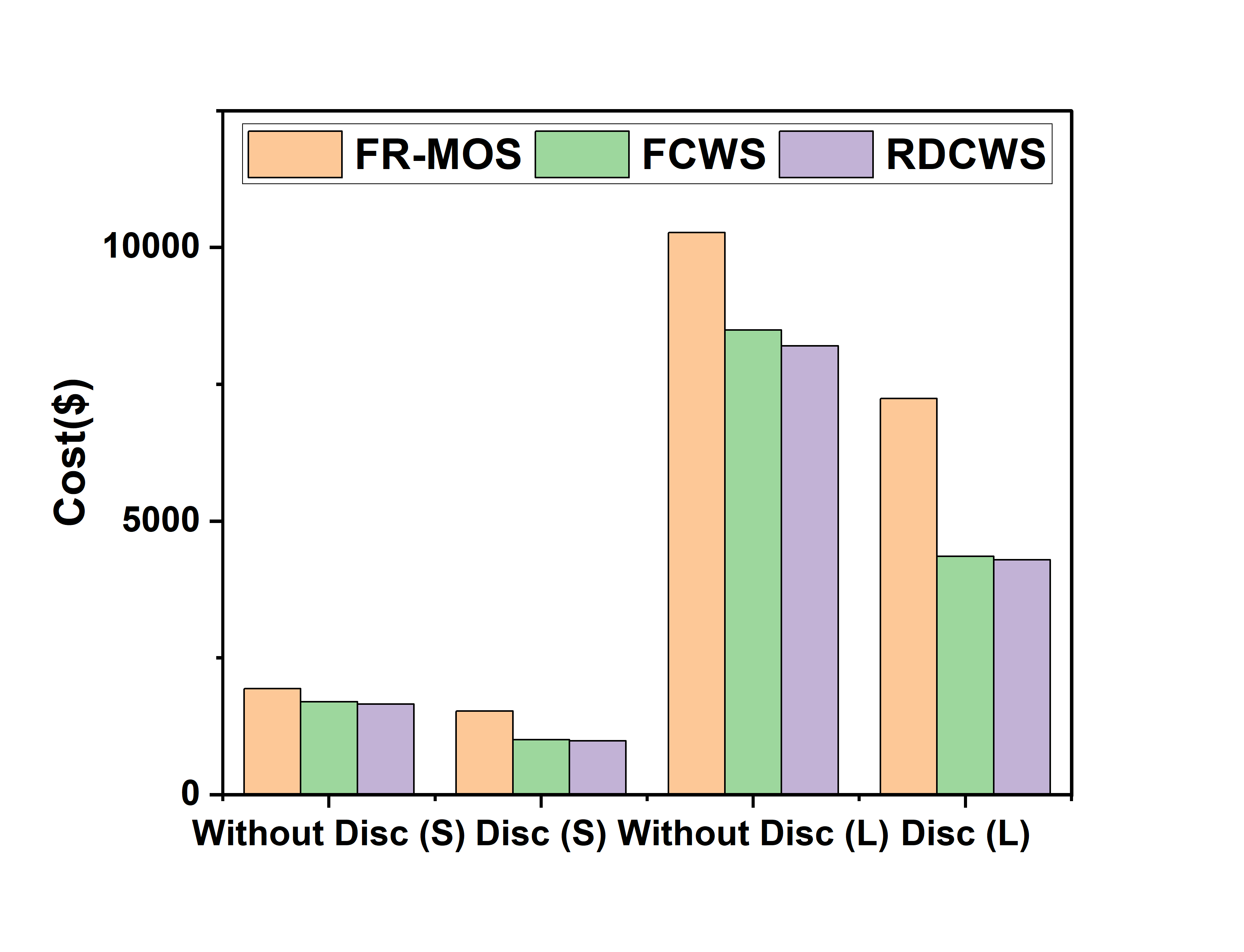} &
\includegraphics[width=0.24\textwidth]{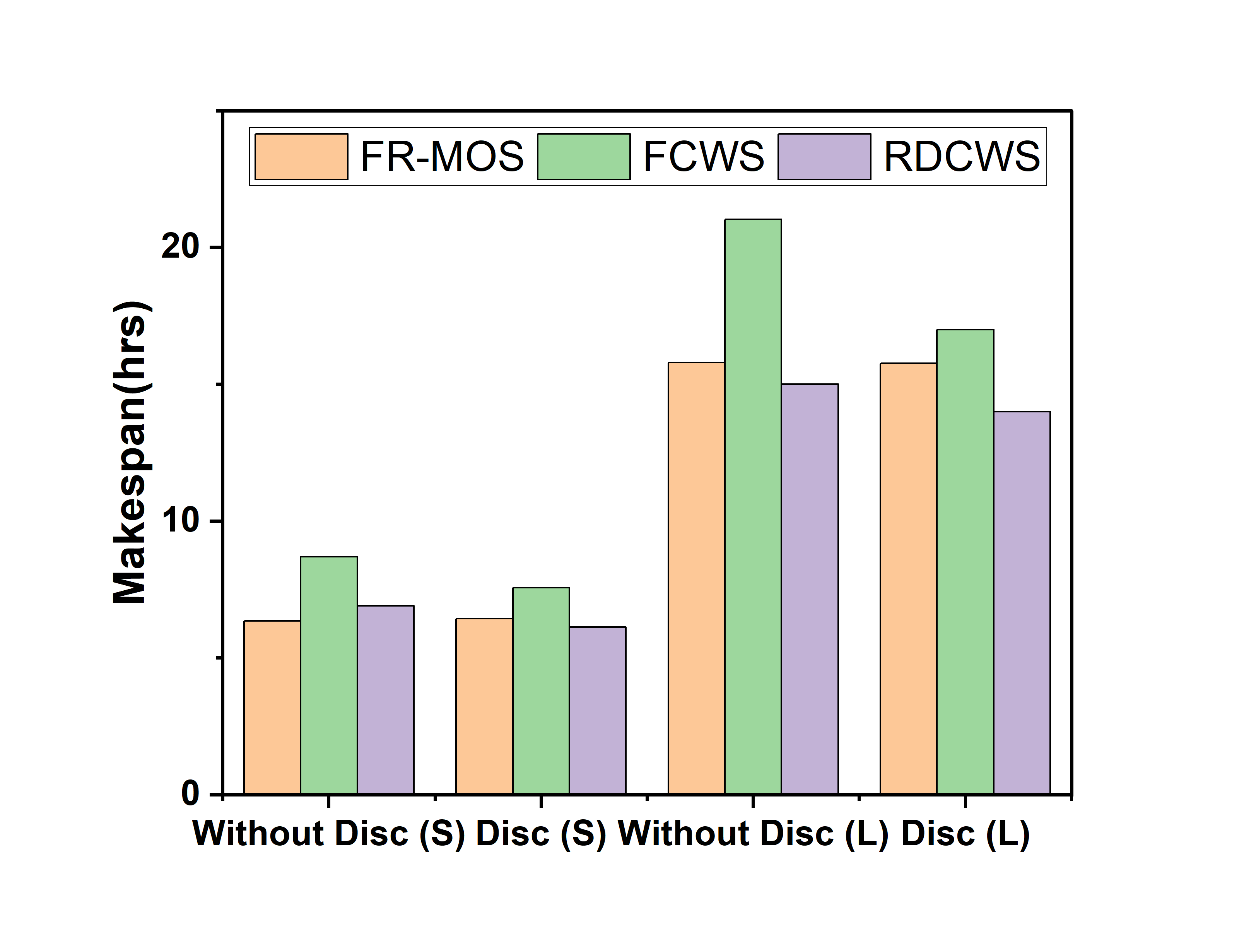} &
\includegraphics[width=0.24\textwidth]{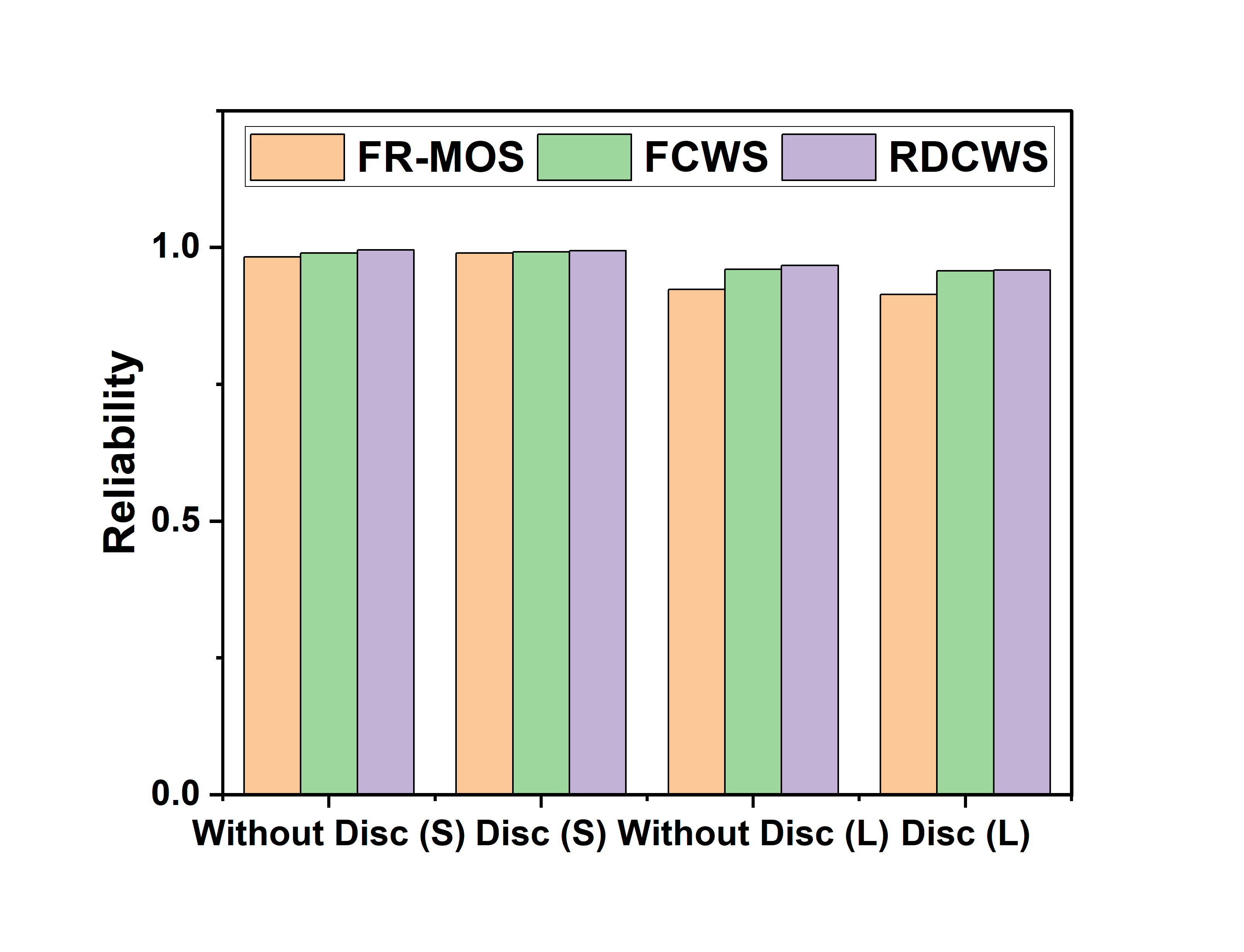}
\\
(a) & (b) & (c) \\
\end{tabular}
\caption{Epigenomics applications with $n = 24, 100$}
\label{Fig4}
\end{figure*}

\begin{figure*}[!ht]
\centering
\begin{tabular}{ccc}
\includegraphics[width=0.24\textwidth]{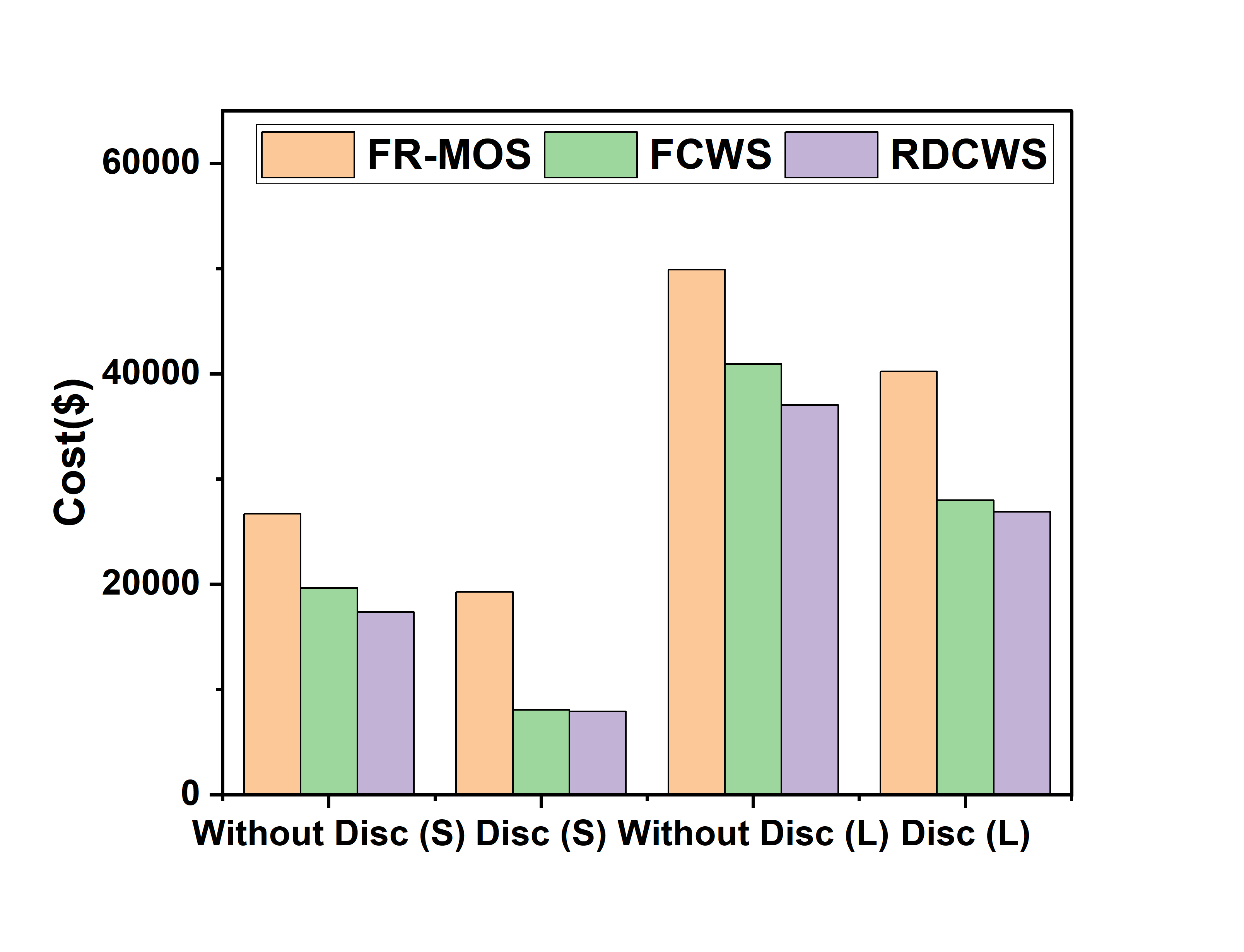} &
\includegraphics[width=0.24\textwidth]{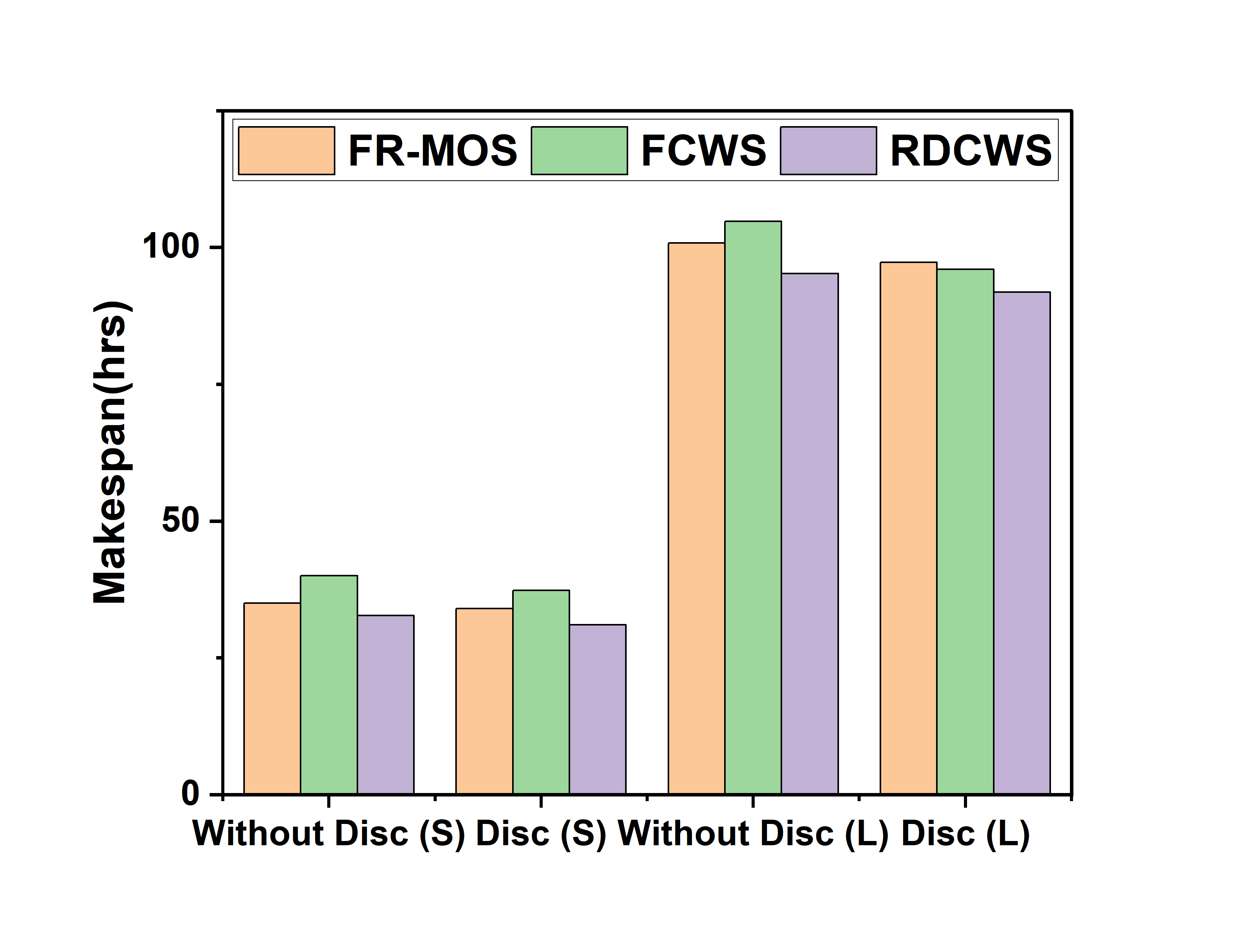} &
\includegraphics[width=0.24\textwidth]{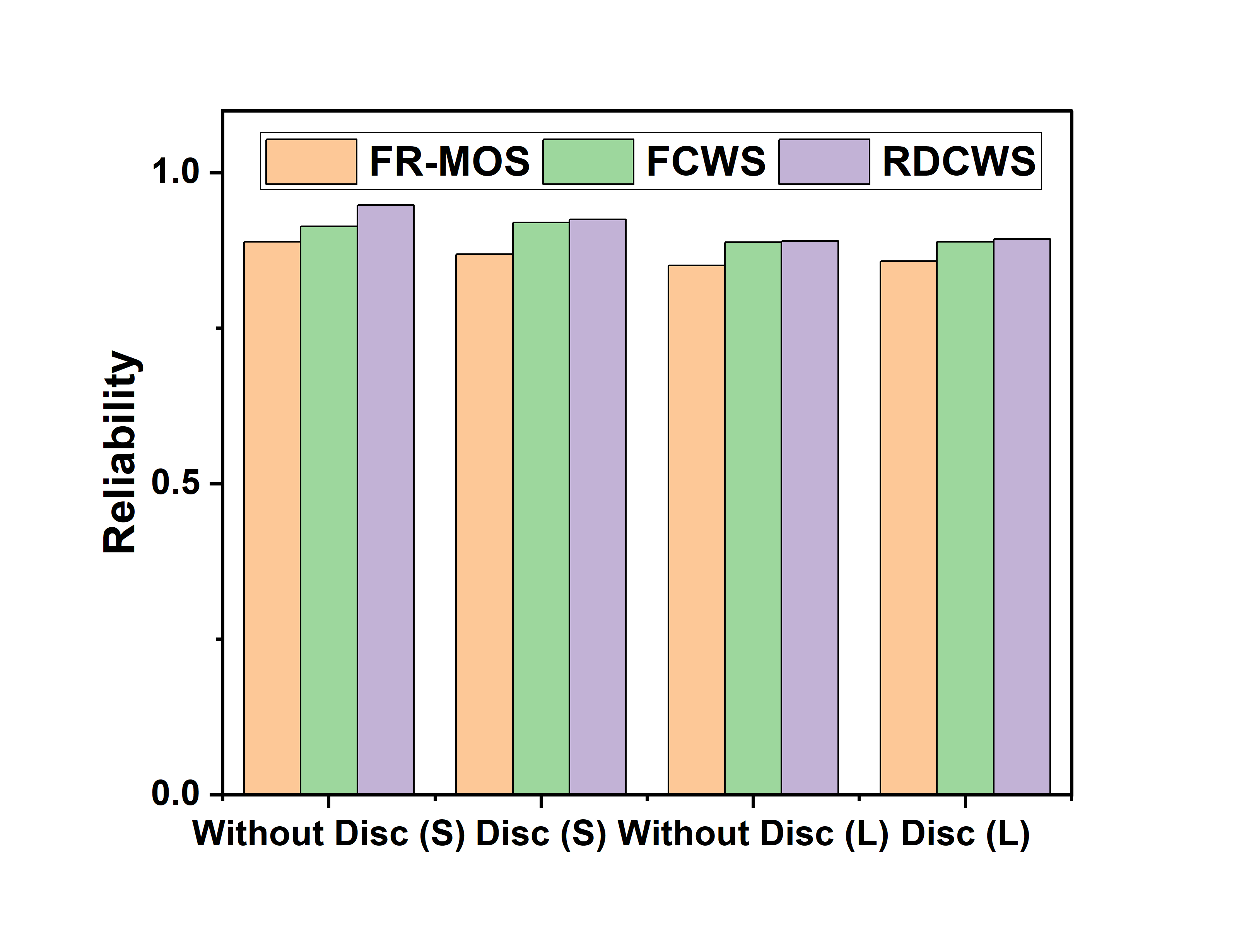}
\\
(a) & (b) & (c) \\
\end{tabular}
\caption{Random applications with $n = 100, 200$}
\label{Fig5}
\end{figure*}

\subsection{Results and Analysis}
Figures \ref{Fig2} - \ref{Fig5} represent the plots for \emph{FFT, LIGO, Epigenomics, Random} workflows respectively. Results below are described in terms of percentages between RDCWS and the best between FR-MOS and FCWS for makespan, cost, and reliability.
\begin{itemize}
\item \emph{FFT:} 
\renewcommand{\theenumi}{\roman{enumi}}%
\begin{enumerate}
	\item \textbf{Cost:} From Fig. \ref{Fig2}(a), we see that RDCWS achieves the least cost with/without considering the discounting scheme. Compared with FCWS for small and large sizes, RDCWS achieves 8.16\%, 8.24\% lesser cost when not considering discounts and 0.45\%, 1.86\% lesser cost when considering discounts, respectively. 
    \item \textbf{Makespan:} From Fig. \ref{Fig2}(b), we see that RDCWS achieves the least makespan with/without considering the discounting scheme. When compared with FR-MOS for small size, RDCWS achieves 5.5\% lesser makespan and outperforms FCWS for large size by 2.36\%. RDCWS outperforms FCWS when discounts are considered by 3.23\% and 3.67\% for small and large sizes respectively.
    \item \textbf{Reliability:} From Fig. \ref{Fig2}(c), we see that RDCWS achieves maximum reliability with/without considering the discounting scheme. When compared with FCWS for small and large sizes, RDCWS achieves 1.14\%, 1.1\% more reliability when not considering discounts and 0.1\%, 0.1\% more reliability when considering discounts respectively.
\end{enumerate}

\item \emph{LIGO:} 
\renewcommand{\theenumi}{\roman{enumi}}%
\begin{enumerate}
	\item \textbf{Cost:} From Fig. \ref{Fig3}(a), we see that RDCWS achieves the least cost with/without considering the discounting scheme. When compared with FCWS for small and large sizes, RDCWS achieves 8.11\%, 5.61\% lesser cost when not considering discounts and 2.93\%, 1.88\% lesser cost when considering discounts respectively.
    \item \textbf{Makespan:} From Fig. \ref{Fig3}(b), we see that RDCWS achieves the least makespan with/without considering the discounting scheme. When compared with FR-MOS for small and large sizes, RDCWS achieves 3.2\%, 0.4\% lesser makespan when not considering discounts. When discounts are considered, it outperforms FCWS by 4.65\%, and 6.28\% for small and large sizes respectively.
    \item \textbf{Reliability:} From Fig. \ref{Fig3}(c), we see that RDCWS achieves maximum reliability with/without considering the discounting scheme. When compared with FCWS for small and large sizes, RDCWS achieves 0.1\%, 1.05\% more reliability when not considering discounts and 0.1\%, 0.1\% more reliability when considering discounts respectively.
\end{enumerate}

\item \emph{Epigenomics:}
\renewcommand{\theenumi}{\roman{enumi}}%
\begin{enumerate}
	\item \textbf{Cost:} From Fig. \ref{Fig4}(a), we see that RDCWS achieves the least cost with/without considering the discounting scheme. When compared with FCWS for small and large sizes, RDCWS achieves 2.65\%, 3.4\% lesser cost when not considering discounts and 2\%, 1.43\% lesser cost when considering discounts respectively.
    \item \textbf{Makespan:} From Fig. \ref{Fig4}(b), we see that when discounts are not considered, for small size, FR-MOS achieves the least makespan giving 8.6\% lesser makespan when compared with RDCWS. For large size, RDCWS outperforms FR-MOS by 5\%. When discounts are considered, RDCWS gives the least makespan outperforming FR-MOS by 4.82\% and 11.17\% for small and large sizes respectively. 
    \item \textbf{Reliability:} From Fig. \ref{Fig4}(c), we see that RDCWS achieves maximum reliability with/without considering the discounting scheme. When compared with FCWS for small and large sizes, RDCWS achieves 1.01\%, 1.04\% more reliability when not considering discounts and 0.1\%, 0.1\% more reliability when considering discounts respectively.
\end{enumerate}

\item \emph{Random:}
\renewcommand{\theenumi}{\roman{enumi}}%
\begin{enumerate}
	\item \textbf{Cost:} From Fig. \ref{Fig5}(a), we see that RDCWS achieves the least cost with/without considering the discounting scheme. Compared with FCWS for small and large sizes, RDCWS achieves 11.58\%, 9.58\% lesser cost when not considering discounts and 1.96\%, 4\% lesser cost when considering discounts, respectively. Hence we see a small improvement for RDCWS over FCWS when discounting scheme is considered compared to when it was not. FCWS gives much lesser cost compared to FR-MOS in all cases. 
    \item \textbf{Makespan:} From Fig. \ref{Fig5}(b), we see that RDCWS achieves the least makespan with/without considering the discounting scheme. When compared with FR-MOS, RDCWS gives 7\%, 5.54\% lesser makespan when discounts are not considered for small and large sizes respectively. The savings are larger when compared with FCWS. When discounts are considered, RDCWS gives 8.82\%, 4.5\% lesser makespan when compared with FCWS for small and large sizes respectively. When RDCWS is compared to FCWS we again observe that the savings in makespan decrease when discounts are considered.
    \item \textbf{Reliability:} From Fig. \ref{Fig5}(c), we see that RDCWS achieves maximum reliability with/without considering the discounting scheme. When compared with FCWS for small and large sizes, RDCWS achieves 3.82\%, 0.01\% more reliability when not considering discounts and 1.09\%, 0.1\% more reliability when considering discounts respectively. When RDCWS is compared to FCWS we again observe a decrease in reliability when discounts are considered.
\end{enumerate}
\end{itemize}

In summary, as the task graph size increases, the makespan and cost increase while the reliability decreases as seen when the plots of small and large size task graphs are compared. For most cases, we see that RDCWS outperforms both FR-MOS and FCWS in terms of all objectives and both scenarios: with and without discounts. In the remaining cases, RDCWS outperforms FCWS in all objectives and is inferior to FR-MOS only in terms of makespan. RDCWS performs significantly better than FCWS in terms of makespan because FCWS considers only cost and reliability while allocation \cite{tang2021reliability}. RDCWS performs significantly better than FR-MOS in terms of cost because it is constrained by fuzzy logic and resource utilization \cite{farid2020scheduling}, \cite{tang2021reliability}.
RDCWS solves these issues by using a normalized linear combination of the concerned metrics depending on the rent time of a task. FCWS always gives lesser cost and higher reliability compared to FR-MOS. All the algorithms result in lesser costs when discounting scheme is considered. As seen from the above comparisons, the solution obtained by RDCWS when discounting scheme is considered gives very small improvement over FCWS compared to when discount scheme is not considered.

\section{Conclusion and Future Research Directions}\label{Sec:CFA}


In this paper, we have studied the problem of scheduling a scientific workflow in a multi-cloud system where the goal is to minimize the makespan, cost and to maximize reliability. We have proposed a complete systems model that integrates communication costs, different billing mechanisms, sustained use discounts and done the task reliability analysis. To solve this problem we have designed a list scheduling approach that allocates tasks based on rent time. In the future we will try to include more practical concerns such as load balance.

\bibliographystyle{IEEEtran}
\bibliography{Paper}

\end{document}